\documentclass[10pt]{article}
\pdfoutput=1
\usepackage{graphicx}
\usepackage[pdftex]{hyperref}
\usepackage{subfigure}
\usepackage[it]{caption}

\begin{document}

\pagestyle{empty}

\begin{center}
{\fontsize{16}{16pt} \bf
 Phase Space Engineering in Optical Microcavities}\\
{\fontsize{14}{14pt} \bf II. Controlling the far-field}\footnote{This
research has been funded in
part by a Strategic Grant from NSERC (Canada) and a Team Project from FQRNT (Qu\'ebec).}\vspace{12pt}\vspace{12pt}\\
{\bf Julien Poirier, Guillaume Painchaud-April, Denis Gagnon, Louis J. Dub\'e\footnote{Corresponding author: ljd@phy.ulaval.ca}}\\
{\it D\'epartement de physique, de g\'enie physique, et d'optique\\ Universit\'e Laval, Qu\'ebec, Qu\'ebec, Canada, G1V 0A6}
\end{center}
\vspace{-0.6cm} \vspace{6pt}
{\bf ABSTRACT}\vspace{0pt}\\
Optical microcavities support Whispering Gallery Modes (WGMs) with a very high quality factor $Q$. However, WGMs
typically display a far-field isotropic emission profile and modifying this far-field profile without spoiling the
associated high $Q$ remains a challenge. Using a 2D annular cavity, we present a procedure capable to achieve these two
apparently conflicting goals. With the correspondence between the classical and the wave picture, properties of the
classical phase space shed some light on the characteristics of the wave dynamics. Specifically, the annular cavity has
a well separated \textit{mixed} phase space, a characteristic that proves to be of crucial importance in the emission
properties of WGMs. While the onset of directionality in the far-field may be achieved through parametric deformation \cite {painchaud_icton10}
of the distance cavity-hole centers, $d$ , this contribution presents a method to
control the emission profile via a second parameter, the hole radius $r_0$. The influence of the classical dynamics to
\textit{control and predict} the field emission will be demonstrated.

\begin{figure}[h]
\subfigure[][]{\includegraphics[width = 4.1cm]{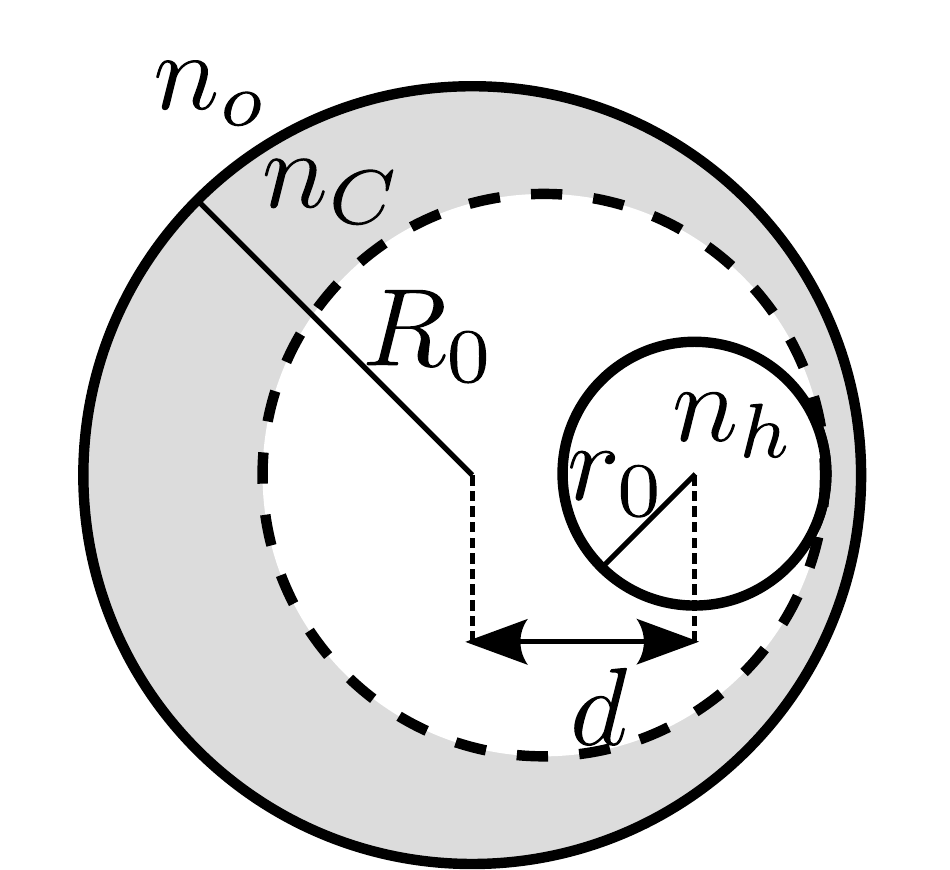} \label{fig1_1a}}
\hfill
\subfigure[][]{\includegraphics[width = 4.1cm]{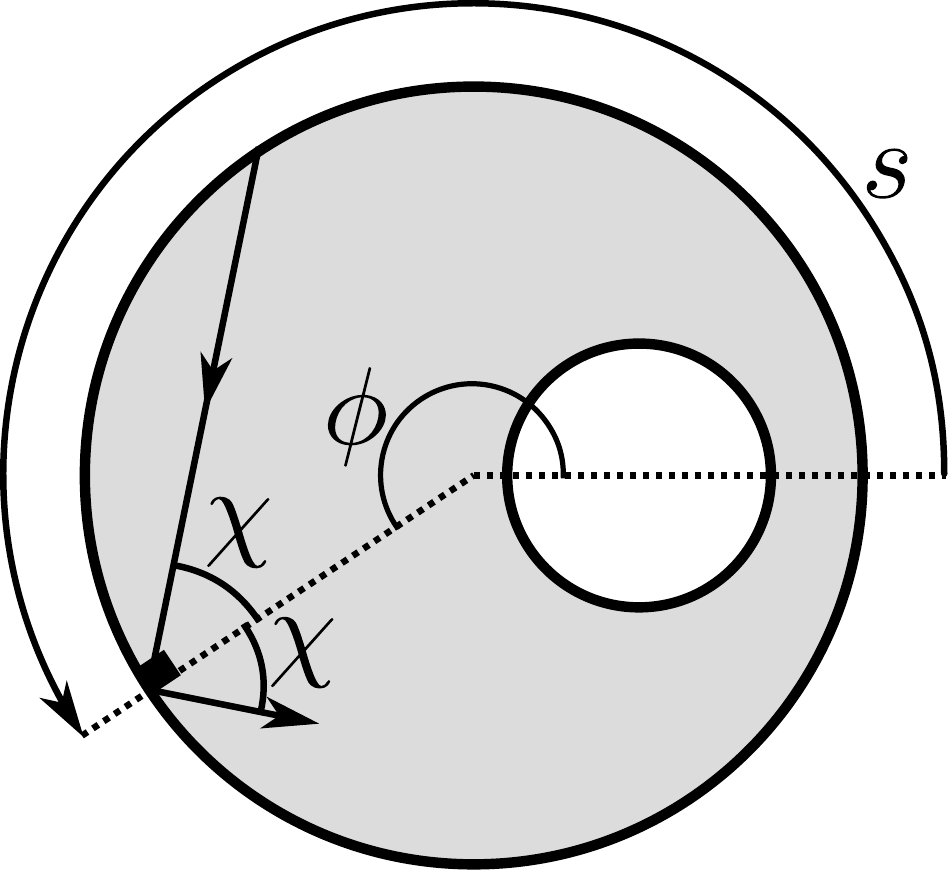} \label{fig1_1b}}
\hfill
\subfigure[][]{\includegraphics[width = 4.5cm]{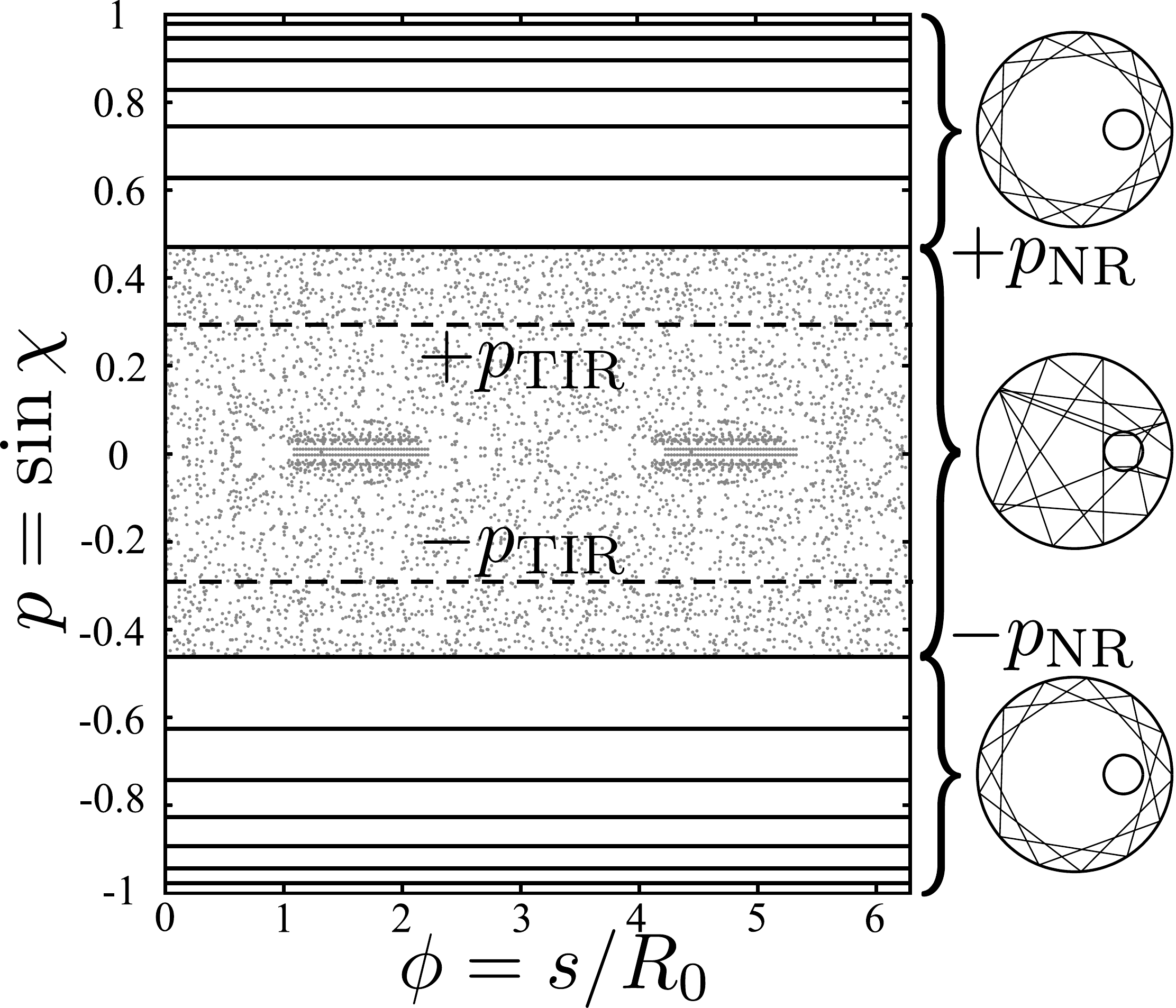} \label{fig1_1c}}

\caption{{\bf Geometry and phase space of the annular cavity.} \textit{
(a) A circular inclusion of refractive index $n_h$ and radius $r_0$ is placed a distance $d$ from the geometrical center of the cavity.
The distance of the inclusion $R_0 - (d+r_0)$ to the boundary will be kept constant in the following; 
(b)  Classical phase space coordinates,  arc length $s$ (alternatively angular position $\phi= s/R_0$) and $p= \sin \chi$; 
(c) Phase space constructed from the impact coordinates $\{\phi_i, p_i\}$ for different initial conditions. Two
distinct regions, containing regular and irregular trajectories respectively, are apparent and well separated by the geometrical condition $|p| = p_\mathrm{NR}=(d+r_0)/R_0$. The boundary of a TIR region ($|p| > p_\mathrm{TIR} \sim 0.3$) is also indicated.}}
\label{fig1_1}
\end{figure}

\noindent{{\bf 1. INTRODUCTION}}\vspace{3pt}\\
Whispering Gallery Modes (WGMs) are known to exist in certain types of microcavities. For instance, the class of regularly shaped dielectric structures includes the sphere, the toroid and the disk (thin cylinder) cavities. The strong confinement of the light field, characteristic of the WGMs, appears as sharp peaks of the spectral response of these dielectric microcavities. Accordingly, these cavities possess high quality (high $Q$) modes. Although this property is of central importance for most, if not all, applications of microcavities, the complete uniformity of the emitted far-field distribution of WGMs remains 
an impediment for useful light source applications (\textit{e.g.} microlasers \cite{noeckel02,vahala03}).
\\
\indent This paper is the sequel of a previous contribution (see \cite{painchaud_icton10} in these Proceedings). We will be interested in the parametrical \emph{control} of the far-field patterns of a WGM located inside an annular dielectric cavity. Our approach follows the observation made in \cite{painchaud_icton10} that the behavior of the far-field of a WGM may be \emph{triggered} from uniformity to non-uniformity using the center-to-center distance $d$ (see figure \ref{fig1_1a}). We will exploit the correspondence between the rise of non-uniformity in the far-field distribution (wave picture) and the extent of non-regular dynamics in classical phase space (ray picture) to predict and control the emission.

\noindent{{\bf 2. CLASSICAL CONCEPTS AND GENERAL CONSIDERATIONS}}\vspace{3pt}\\
\noindent The geometry of the annular dielectric cavity \cite{hentschel02b}
 is depicted in figure \ref{fig1_1a}. It consists of an outer disk of radius $R_0$ and
refractive index $n_C$ in which a circular inclusion of radius $r_0$ and refractive index $n_h$ is embedded and whose center is a distance
$d$ from the center of the main disk.  The cavity is surrounded by a dielectric medium of index $n_o$. In the classical limit, the dynamics within
the cavity amounts to follow individual rays as they propagate freely in between collisions with the outer boundary and the inclusion. 
For an ensemble of many trajectories, the phase space is constructed from the successive coordinates
$\{s_{i},p_{i}\}$ (figure \ref{fig1_1b}) recorded on the surface of the outermost boundary: $\{s_i\}$ are the arc length 
and $\{p_i= \sin \chi_i\}$ where $\chi_i$ is the angle incidence with respect to the normal. Note that the rays are specularly reflected at the outermost
boundary whereas we allow transmission ($|p| < n_h/n_C$) or reflection ($|p| > n_h/n_C$) at the inclusion interfaces. The result
is displayed in figure (\ref{fig1_1c}). Several features are worth mentionning, some of them unique to the annular cavity.\\
\indent The system under study possess a {\em mixed} dynamics. Even more, phase space is well divided into {\em two} principal regions: 
a {\em regular} domain,   $|p| > p_\mathrm{NR} \equiv (d+r_0)/R_0$ ,
filled with trajectories, reminiscent of the integrable disk dynamics, oblivious to the inclusion, and a {\em non-regular} domain,  
$|p| < p_\mathrm{NR}$ where irregular, mostly chaotic, trajectories are observed. The extent of the region where the dynamics is non-regular
is simply governed by the position $d$ and size $r_0$ of the inclusion and proportional to $d+r_0$. 
Another domain of interest is the classical {\em emission} (or escape) region defined by the 
$E=\{(s,p)\ : 0 \leq s \leq 2\pi R_0, |p| \leq p_\mathrm{TIR}\}$, under the Snell-Descartes condition of Total Internal Refexion (TIR) at the outer boundary
\begin{equation}
    p_\mathrm{TIR} = n_o/n_C\quad . \label{eq1_3}
\end{equation}
Clearly for $p_\mathrm{NR} > p_\mathrm{TIR}$, the classical far-field emission is strictly determined by the non-regular ``chaotic'' trajectories \cite{painchaud_icton10}. Moreover, only a small subset of $E$ is actually accessible for emission, we call this subset the {\em effective} emission region,
and denoted it by $W$.  It consists of the trajectories initially located outside of $E$ that get mapped to $E$ through a single iteration of the (implicit) Poincar\'e map, ${\cal P}: \ (s_i,p_i)\mapsto(s_{i+1},p_{i+1})$ . The region $W$ is then defined as
\begin{equation}
    W = {\cal P}({\cal P}^{-1}(E)\cap \bar{E})\label{eq1_5}
\end{equation}
with the complement $\bar{E}=\{(s,p)\ : 0 \leq s \leq 2\pi R_0,\ p_{\mathrm{TIR}}\leq|p|<1\}$. Trajectories entering $E$ are thereafter bound to escape the cavity. The dynamical system so-defined does not conserve energy (\textit{i.e.} the total amount of initial conditions (or initial intensity) decreases as the trajectories reach $E$).
\\
\indent The domain $W$ is obviously modified by parametric variations of both $d$ and $r_0$.  Since we have observed that the far-field emission of a WGM becomes \emph{non-uniform} with increasing $d$ (for fixed $r_0$) \cite{painchaud_icton10,painchaud10}, and accordingly increasing $p_\mathrm{NR}$, we propose that 
the \emph{directionality} of the emission  may be further engineered by keeping $d+r_0$ ($p_\mathrm{NR}$) constant and changing $r_0$,
thereby modifying the ``mixing'' properties of the non-regular region without affecting its size. The quantitative modifications 
induced  will now be investigated in the wave dynamics under the guidance of the previous classical considerations.

\begin{figure}
  \centering
  \subfigure[\label{fig2_1a}]{\includegraphics[height=0.25\textwidth]{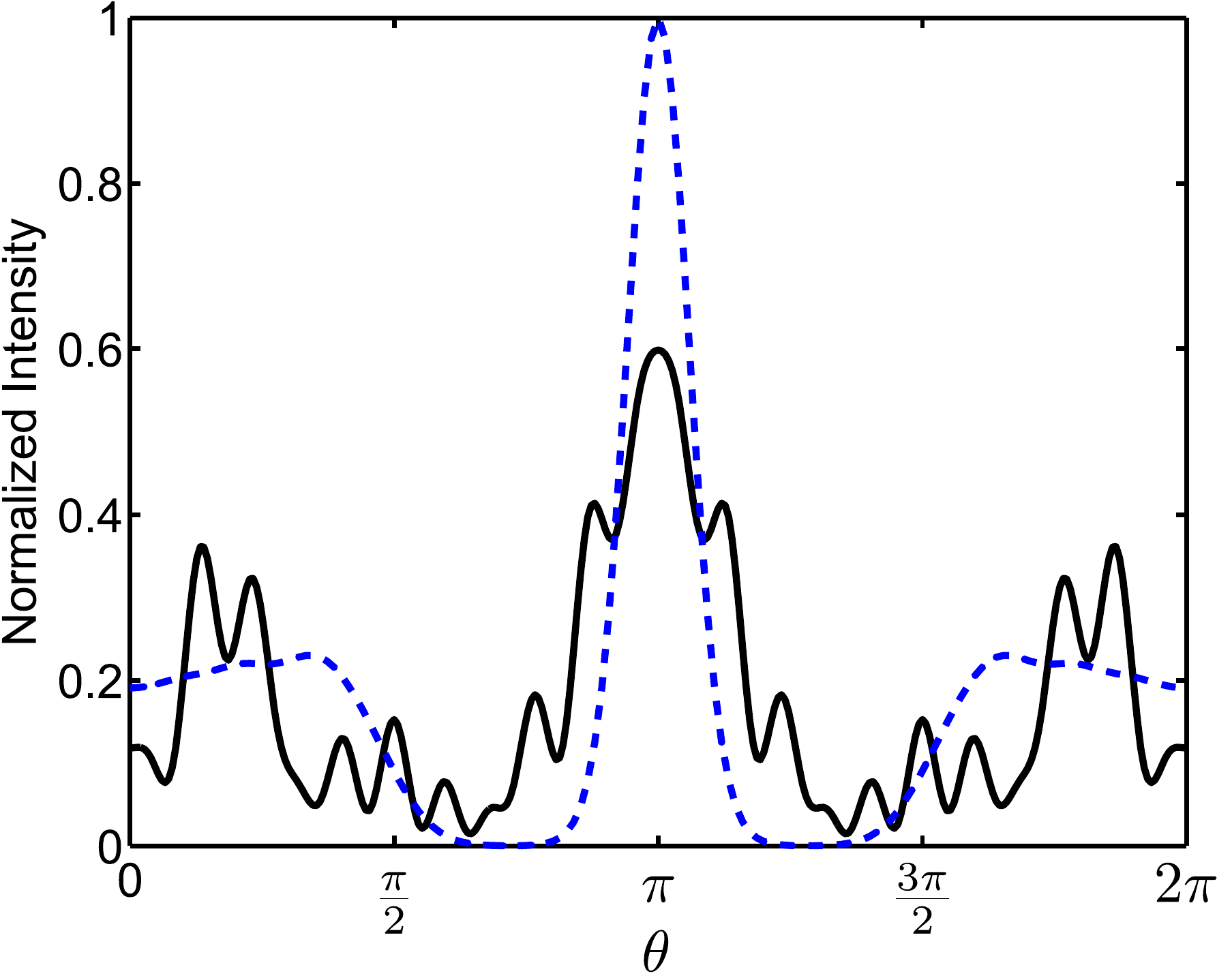}}\hspace{0.5cm}
  \subfigure[\label{fig2_1b}]{\includegraphics[height=0.25\textwidth]{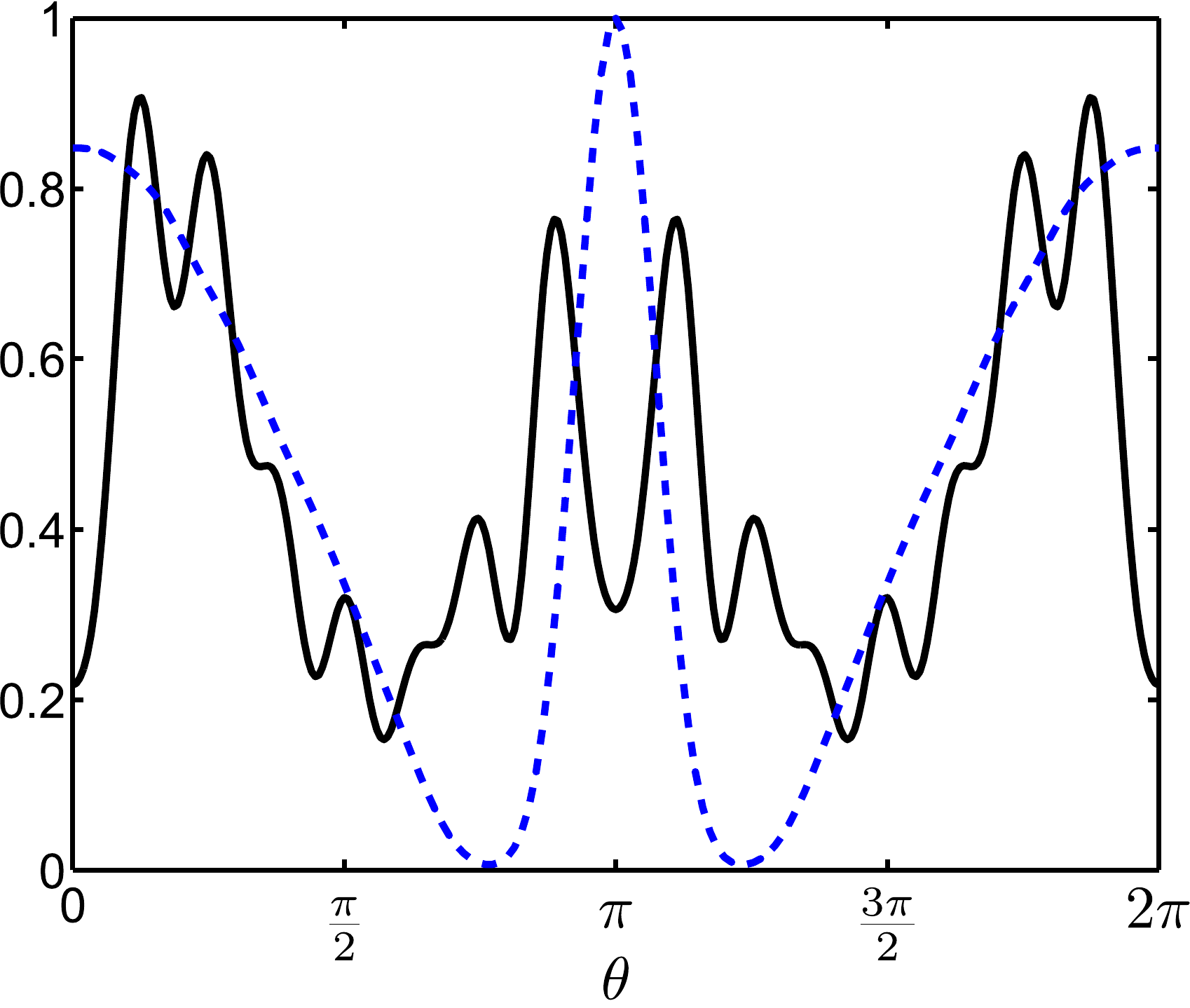}}\hspace{0.5cm}
  \subfigure[\label{fig2_1c}]{\includegraphics[height=0.25\textwidth]{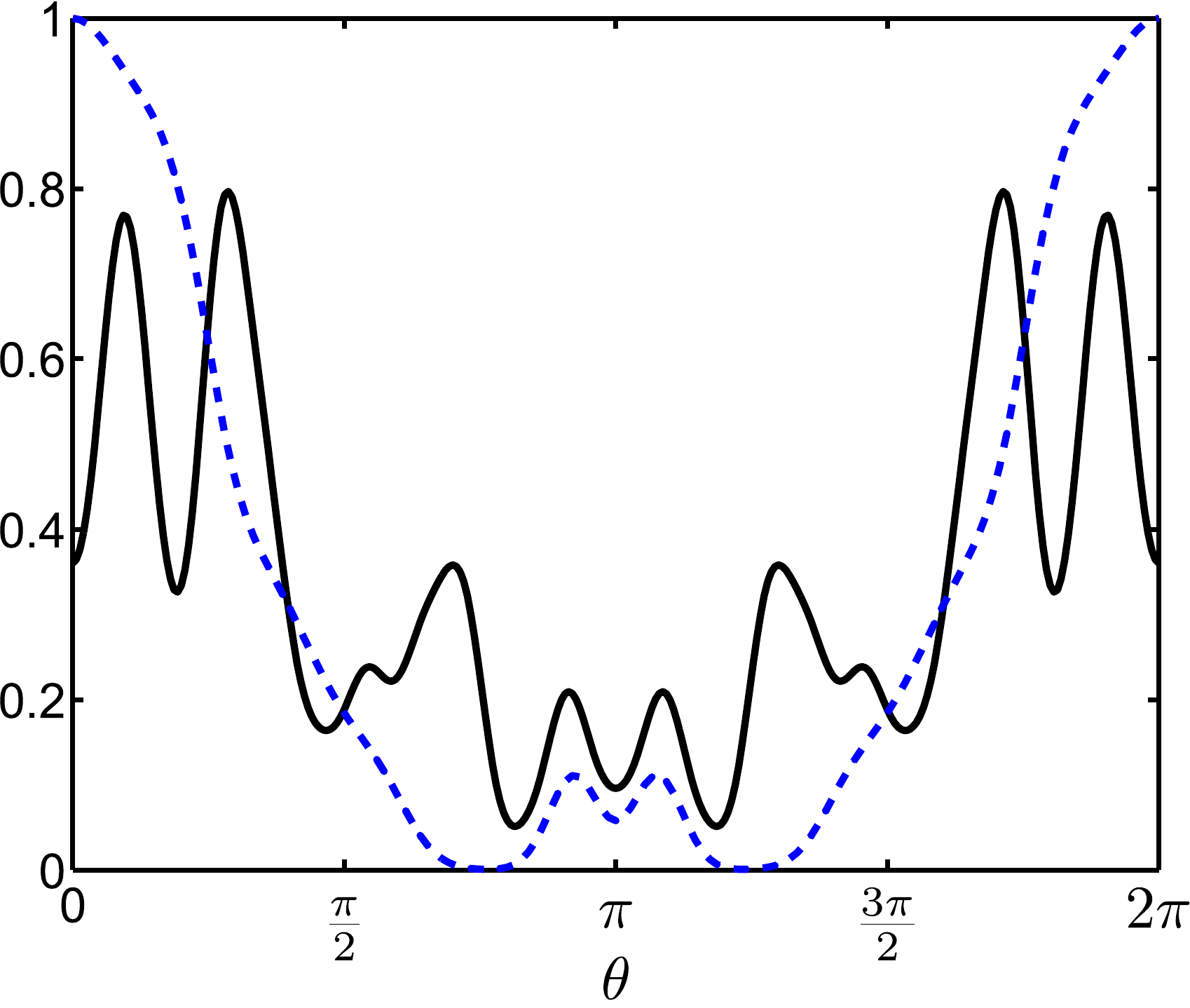}}\\
  \caption{{\bf Far-field intensities.} \textit{ full curves: combined envelope of the two symmetries of mode $(11,1)$, $|\psi_{(11,1)}^e|^2+|\psi_{(11,1)}^o|^2$; dashed curves: non-coherent sum of the classical trajectories reaching the far-field
for 3 values of $r_0$, (a) $r_0= 0.064 R_0$, (b) $r_0= 0.127 R_0$, (c) $r_0= 0.211 R_0$, while $d+r_0$ is kept fixed at $0.55 R_0$. 
For ease of comparison, the intensities in each panel have been arbitrarily normalized to the maximum of the classical intensities.}
 \label{fig2_1}}
\end{figure}

\noindent{{\bf 3. EFFECT OF PARAMETER $r_0$ ON THE FAR-FIELD BEHAVIOR}}\vspace{3pt}\\
We approximate our physical annular cavity by a planar (2D) system and search for resonant modes (here TM polarization) 
by solving Helmholtz equation
\begin{equation}
    \left[\nabla^2 + n^2(r,\phi)k^2\right]\psi(r,\phi) = 0 \label{eq3_1}
\end{equation}
under appropriate boundary conditions. The solutions are obtained by a careful implementation of the scattering formalism of \cite{rahachou04a}. The physical parameters of the system are set as $n_C = 3.2$, $n_h=n_o=1$, $R_0=1$ keeping the control group  $d+r_0$ constant at = 0.55 $R_0$. 
The important regions are then delimited by $p_\mathrm{TIR}= 0.3125$ and $p_\mathrm{NR}= 0.55$. 
For the sake of demonstration, we have selected the high $Q$ WGM $(11,1)$ at  $k R_0\approx 4.5$. The notation $(m,n)$ is meant to identify
the  angular momentum number and the number of radial nodes respectively, and $k$ is the wavenumber. The particular choice of $d+r_0$ is motivated by the high far-field contrast value $C_{11}$ for $r_0=0.20 R_0$ and $d=0.35 R_0$ \cite{painchaud_icton10}. 
This contrast measure $C_{m_0}$ varies between 0 and 1: 0 meaning that the
observed far-field has fully retained its original uniform $m_0$ character, and 1 indicating that the perturbed mode has  mixed
with other angular momentum components and that its field distribution is highly non-uniform.    
We have retained the two symmetries of mode $(11,1)$ in the full-wave calculations (even/odd, $\psi_{(11,1)}^{e/o}$, relative to the center-to-center axis).
\\
\indent We follow the mode $(11,1)$ as  $r_0$ is varied from $0.05 R_0$ to $0.40 R_0$ and although its quality factor drops from about $10^7$ to roughly $10^5$ over the interval, it remains the dominant resonance, albeit with a changing character. The far-field emission is recorded as
a function of the observation angle $\theta$ from 
the non-coherent sum of $|\psi_{(11,1)}^e|^2$ and $|\psi_{(11,1)}^o|^2$. 
The far-field intensities are displayed in figure (\ref{fig2_1}) for 3 separate values of $r_0$. The emission profiles evolve from
a sharp contribution near $\theta=\pi$ at $r_0 = 0.064 R_0$ (figure \ref{fig2_1a}) to an increasingly broad multi-peaks envelope 
with equally important contributions around $\theta= \pi/4, \pi$ and $3\pi/4$ at $r_0= 0.127 R_0$ (figure \ref{fig2_1b}) and
to a depleted emission in the backward direction in favor of two important forward peaks at $r_0= 0.211 R_0$ (figure \ref{fig2_1c}).
In short, the objective of modifying the directionality of the far-field emission of a WGM through parametric control is achieved.
\\
\indent To connect with the intuition gained in the previous section, we have carried out a ray-escape simulation \cite{schwefel04}
in the classical representation. Assuming that the transition probability to the effective escape region $W$ is highest near the $p_\mathrm{NR}$ limit,
initial conditions are chosen uniformly in a thin strip of phase space with $p \sim p_\mathrm{NR}$. For a given $r_0$, each initial condition $(s_0,p_0)$
is followed through consecutive impacts $(s_i,p_i)$ with the outer boundary where a reflected intensity $I_{i+1}= I_i (1 -|T(p_{i+1})|^2)$ is calculated
 with Fresnel transmission coefficient,
\begin{equation}
    T(p) = \frac{\sqrt{1-p^2}}{\sqrt{1-p^2}+\sqrt{(n_o/n_C)^2-p^2}}\quad . \label{eq3_2}
\end{equation}
Iterations stop when the intensity associated with a trajectory has dropped from its initial value $I_0$ to an arbitrarily small number. The escaping intensities of every trajectory are then binned according to observation angles $\theta$ to form the far-field distribution.
 Finally, the classical far-field is convoluted with a gaussian function of width $\sigma=2\pi/4m_0$ to smooth out irregularities. 
The width is chosen to account for the angular resolution of the pure WGM $(m_0,n)$ which has $2 m_0$ lobes in the azimuthal direction.
\\
\indent The comparison with the full-wave far-field profiles is presented in figure \ref{fig2_1}. At this low value of $n_CkR_0 \sim 15$, it is
no surprise that the agreement is only qualitative although the main features are present in both calculations. Our exploratory  simulations 
are nevertheless  encouraging and we expect the procedure to deliver better correspondence at higher values of $n_CkR_0$. One should stress 
however that the classical approach is universal and generic. In the annular geometry studied here, a large number of high $Q$ WGM $(m,n)$
are available above $|p|= p_\mathrm{NR}$ and they are well localized around their resonance condition  $p_\mathrm{WGM}= m /(n_C k R_0)$.
As long as $p_\mathrm{WGM} > p_\mathrm{NR}$, they are candidate for far-field directionality while preserving 
their near-field uniformity, i.e. low-loss, high $Q$ character, and phase space design is a possible route to optimization.
\\
\begin{figure}[h]
  \centering
  \includegraphics[width=0.4\textwidth]{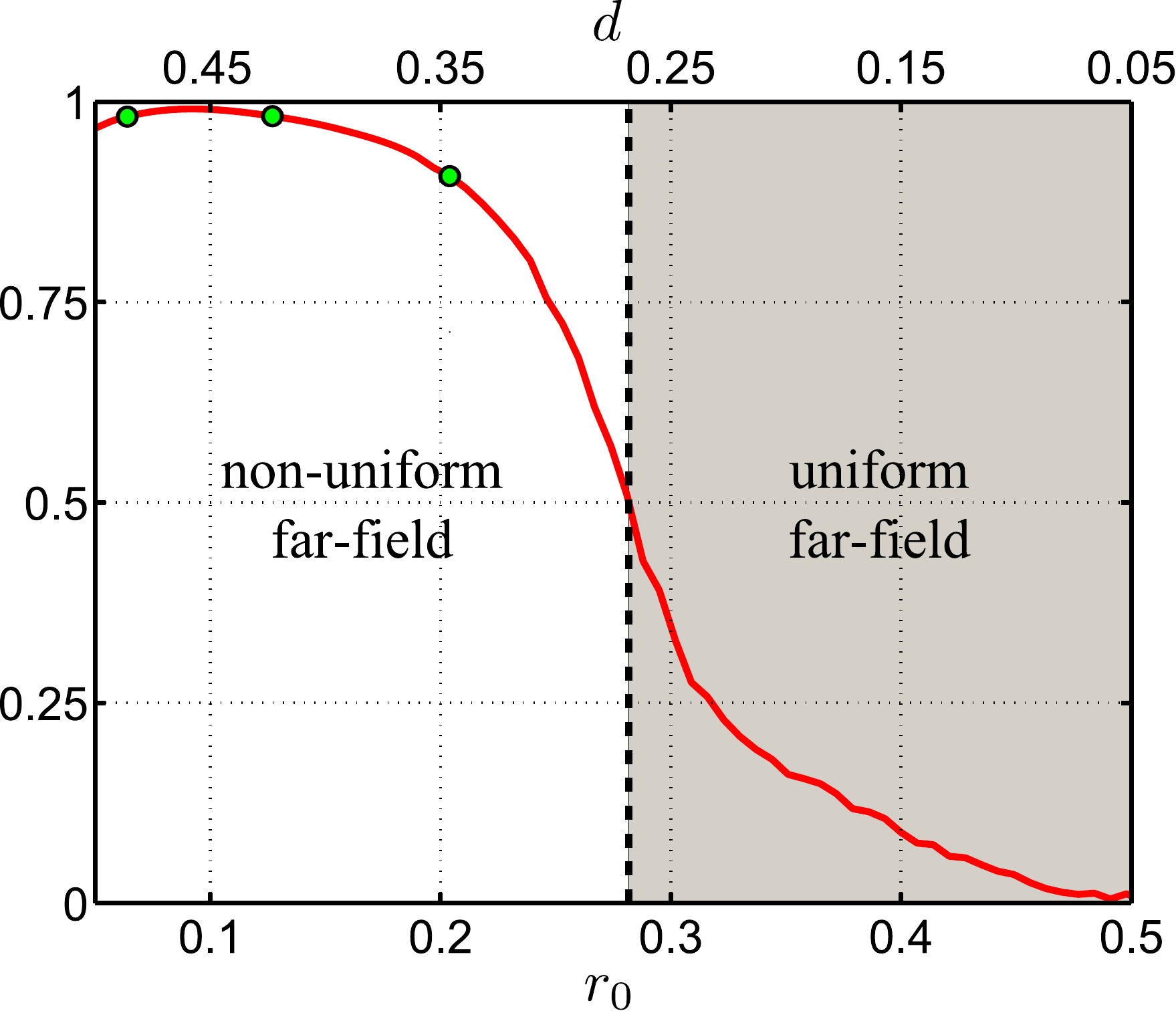}\\
  \caption{{\bf Constrat measure $C_{11}$ versus $r_0$.} \textit{For fixed $d+r_0 = 0.55 $ (distances in unit of $R_0$),
2 regions, non-uniform and uniform far-field, arbitrarily divided at $C_{11} = 0.5$, define the range of $r_0$ for which
the  far-field profile has a certain character. The 3 dots correspond to the same values of $r_0$ presented in figure (2).} \label{fig3_1}}
\end{figure}
\indent Figure (\ref{fig3_1}) gives a sense of the range of values of the control parameters where one can modify (engineer) 
at will the far-field distribution keeping a high  quality factor. The contrast measure $C_{11}(r_0)$ stays above 0.80, meaning strong mixing
 with other angular components $m \not= m_0$, for $r_0 \in [\sim 0, \sim 0.25]$. As $r_0$ is further increased, $C_{11}$ drops further until it 
reaches low values where the WGM has recovered its regular uniform character.

\noindent{{\bf 4. CONCLUSION}} \vspace{3pt}\\
In this paper we have shown that it is possible to control the directionality of the far-field emission through the modification of the radius $r_0$ of
a circular inclusion in a annular dielectric cavity. Evidences have been presented that the far-field profile of both full-wave and classical simulations 
show similar structures. Furthermore, for a given annular geometry, the classical far-field profile is universal. Our results along with those of \cite{painchaud_icton10,painchaud10} pave the way to design scenarios for high $Q$ directional emission from
WGMs of disk-shaped microcavities.
\\

\newpage
\noindent{{\bf REFERENCES}}\vspace{3pt}\\

\end{document}